\documentclass[twocolumn,aps,pra,showpacs]{revtex4}%
\usepackage{amsmath}%
\usepackage{amsfonts}%
\usepackage{amssymb}%
\usepackage{graphicx}
\usepackage{epstopdf}
\usepackage{xcolor}

\begin{document}

\title{Preventing side-channel effects in continuous-variable quantum key distribution}
\author{Ivan Derkach}
\email{ivan.derkach01@upol.cz}
\affiliation{Department of Optics, Palack\'y University, 17. listopadu 12, 77146 Olomouc, Czech Republic}
\author{Vladyslav C. Usenko}
\email{usenko@optics.upol.cz}
\affiliation{Department of Optics, Palack\'y University, 17. listopadu 12, 77146 Olomouc, Czech Republic} 
\author{Radim Filip} 
\email{filip@optics.upol.cz}
\affiliation{Department of Optics, Palack\'y University, 17. listopadu 12, 77146 Olomouc, Czech Republic}
\date{\today}
\begin{abstract}
The role of the side channels in the continuous-variable quantum key distribution is studied. It is shown
how the information leakage through a side channel from the trusted sender station increases the vulnerability of the protocols to the
eavesdropping in the main quantum communication channel. Moreover, the untrusted noise infusion by an eavesdropper on the trusted receiving side breaks the security even for a purely attenuating main quantum channel. As a method to compensate for the effect of the side-channel leakage on the sender side, we suggest several types of manipulations on the side-channel input. It is shown that by applying the modulated coherent light on the input of the side channel that is optimally correlated to the modulation on the main signal and optionally, introducing additional squeezing in the case of the squeezed-state protocol, the negative influence of the lossy side channel on the sender side can be completely removed. For the trusted receiving side, the method of optimal monitoring of the residual noise from the side-channel noise infusion is suggested and shown to be able to completely eliminate the presence of the noisy side channel. We therefore prove that the side-channel effects can be completely removed using feasible operations if the trusted parties access the respective parts of the side channels.
\end{abstract}
\maketitle

\section{Introduction}
Quantum key distribution (QKD) \cite{BB,QKDRev} is a major communication application
of quantum information theory aiming at the development of protocols
for establishing secure channels protected by the laws of quantum physics.
Such channels can then be used to share a secure key for classical 
symmetrical cryptographic systems. Recently, continuous-variable (CV) \cite{Cm}
protocols of QKD (see \cite{Diamanti2015} for review) were developed and implemented on the basis of squeezed \cite{Ralph1999,Cerf2001,Madsen12} or coherent \cite{Grosshans2002,Grosshans2003a,Silberhorn2002,Jouguet2013,Huang2016} states.
The security of CV QKD protocols in the case of Gaussian modulation was then shown against collective attacks in the presence of channel noise \cite{coll1, coll2}, which also implies the security against the most general coherent attacks \cite{Renner07, Mertz13}.

CV QKD protocols, however, suffer from various imperfections. The most threatening are the untrusted (i.e., being under full control of a potential eavesdropper) quantum channels, which are inclined to losses due to the attenuation and can add excess noise in the link. Such noise can also be detection noise indistinguishable from the effect of the channel. In security analysis it is then supposed that all the channel imperfections are due to the presence on an eavesdropper. It was an important step in the development of CV QKD when with the use of reverse reconciliation it was shown possible to establish asymptotically secure key transmission upon any pure channel loss \cite{Grosshans2003a}, while noise remains limiting to the security of the protocols. 

However, the insecure quantum channel is not necessarily the single source of information leakage from a QKD protocol. A potential eavesdropper can use imperfections of the trusted (i.e., fully controlled by the trusted parties) devices such as sources and detectors to gain at least partial information on the signal being sent or to control the measurement being performed at the receiver station. The noise, which is present on the trusted sides, can be fully controlled and calibrated by the trusted parties. Such noise, however, can still be harmful. It was shown in particular that the preparation noise can already break the security in the reverse reconciliation protocol \cite{Filip08}, but can be suppressed \cite{Usenko10} or tolerated in the direct reconciliation scheme \cite{QKDClassLim,QKDClassLim2}. Also, the trusted detection noise limits the key rate, but can be partially helpful to make the protocol more robust against noise in the quantum channel \cite{Cerf,Usenko2016}. 

In the less optimistic scenario the noise or loss on the trusted sides can however be under partial control of an eavesdropper, as depicted in Fig. \ref{scheme}(a). This is the case of the side channels, which we define as auxiliary channels that have either input or output controlled by a trusted party but output or input, respectively, controlled by an eavesdropper. From this point of view, the side channels differ from the main channel between the sender and receiver. Supposedly, any additional information can be used by an attacker to increase the knowledge about the transmitted key. Therefore, it is necessary to investigate the influence the side channels can have on security. In the following study we summarize all possible sources of side information and define them together as the side channels on either the sender or the receiver side of the protocol.



One possible way to overcome the negative influence of the side channels is implementing the so-called measurement-device-independent (MDI) QKD protocols \cite{deviceind}, which were recently suggested on the basis of CVs \cite{deviceindCV,deviceindCV2}, where the trusted detection stations become shielded from a potential eavesdropper. However, the applicability of the device-independent CV QKD protocols is still very limited, particularly in terms of distance. 

In the present paper we study the effect of the side channels in CV QKD protocols with coherent and squeezed states of light. We define the side channels as the imperfections (signal loss and noise) on the trusted sides, which are under partial control of an eavesdropper. In particular, we consider (A) the leakage from the trusted sender station and (B) measurement manipulation by the noise addition in the trusted receiver station.  We show the degradation of the key rate and increase of vulnerability to the channel noise in the presence of a side-channel leakage. We also show a security break from the noisy side channel on the detection stage. We suggest methods to compensate for the negative influence of the described side channels. For (A) we consider the possibility to classically apply an additional correlated signal on the side-channel input, which is under control of a trusted sender party. We show the positive effect of such additional modulation and the possibility to optimize the modulation variance for the given parameters of the protocol. Moreover, we show that by applying correlated information encoding and squeezing the input of the side channel, in the case of the squeezed-state protocol, the trusted party is able to completely decouple the side channel from the signal. By decoupling here we mean decorrelation (reducing or turning the correlation to zero) and stopping the leakage of information through the side channel, which completely removes the negative impact of the side channel. For (B) we show the possibility to cancel the infused detection noise by monitoring the output of such a noise-infusing side channel. These are the alternative ways of active compensation of the side channels in the Gaussian CV QKD protocols with the trusted sender and receiver stations, which keep the advantage of usability of such protocols, including the longer channel distances, compared to the device-independent protocols \cite{deviceind, deviceindCV,deviceindCV2}, and do not involve entanglement or non-Gaussian operations and measurements. If for any reason the input of the sender-side leakage or the output of the receiver-side noise infusion are not available for the manipulations or monitoring, respectively, then the negative impact of the side channels shown in the current paper has to be either taken into account in the security analysis or compensated for by the possible use of the MDI schemes.

The paper is structured as follows. In Sec. \ref{security} we define the side channels and recapitulate the methods of CV QKD security analysis being used. In Sec. \ref{prob} we demonstrate the negative impact of the side channels on the CV QKD security. In Sec. \ref{solve} we introduce the methods aimed at compensating for the negative effect of the side channels. We summarize in Sec. \ref{concl}.

\section{Types of side channels.}
\label{security}
\begin{figure}
	\centering
		\includegraphics[scale=0.55]{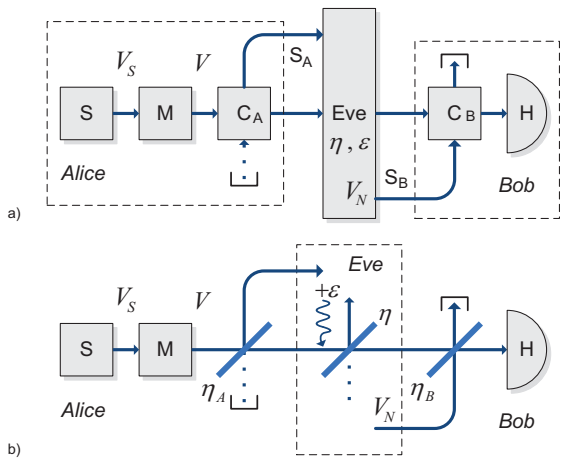}
	\caption{(a) Scheme of the CV QKD based on signal state preparation in the source $S$ and Gaussian quadrature displacement applied in the modulator $M$. The untrusted channel is parametrized by transmittance $\eta$ and excess noise $\epsilon$. The signal is coupled to the lossy side channel $S_A$ with untrusted output on the sender side and to the noisy side channel $S_B$ with untrusted input of variance $V_N$ on the receiving side. The remote trusted party performs measurement with the homodyne detector H. The {\em trusted} devices and channels are within the dashed boxes. (b) Scheme of the CV QKD with sender-side leakage modeled as coupling of the signal to a vacuum mode on a beam splitter with transmittance $\eta_A$. The receiver-side untrusted noise infusion is modeled as coupling to a noisy mode with variance $V_N$ on a beam splitter with transmittance $\eta_B$. The {\em untrusted} channels are within the dashed box. }
	\label{scheme}
\end{figure}

We study the effect of the side channels on the standard and optimized CV QKD protocols \cite{Madsen12,GaussRev,Usenko11} on the basis of the Gaussian modulation of squeezed and coherent states, as depicted in Fig. \ref{scheme}(a). The trusted sending side (Alice) prepares the signal state (squeezed or coherent) with variance $V_S$ (so that $V_S<1$ or $V_S=1$, respectively) using the source $S$. Alice then applies random Gaussian quadrature displacement of variance $V_M$ (also referred to as the modulation variance), so that the overall variance of the modulated states becomes $V$, using the modulator $M$. The prepared state travels through the untrusted channel parametrized by transmittance (loss) $\eta$ and excess noise $\epsilon$ both being under {\em full} control of an eavesdropper (Eve). The signal is then detected by the remote receiving party (Bob) using the homodyne detector H.  Further, with no loss of generality, we assume that the quadrature $x$ is measured by Bob. Thus, in the standard Gaussian CV QKD protocol (without the side channels) Alice applies displacement $x_M$ to the signal quadrature $x_S$ and sends the state with the quadrature $x_A=x_S+x_M$ to the channel so that the variances are $Var(x_A)=V$, $Var(x_S)=V_S$ and $Var(x_M)=V_M$ and then $V_S+V_M=V$. In the standard Gaussian CV QKD squeezed-state protocol \cite{GaussRev} the signal states are modulated up to the antisqueezing (variance of the quadrature complementary to a squeezed one), so $V_M=1/V_S-V_S$ holds, i.e., the variance of modulation is fixed by squeezing of the signal states. We will also consider the optimized Gaussian CV QKD protocols \cite{Usenko11,Madsen12}, where modulation $V_M$ is independent of the variance of the signal states and can be freely optimized for a given signal resource and parameters of the setup. The quantum channel transforms the modulated signal such that Bob measures the quadrature $x_B=(x_A+x_N)\sqrt{\eta}+x_0\sqrt{1-\eta}$, where $x_0$ is the quadrature of the vacuum input of the channel loss and $x_N$ is the quadrature of the channel excess noise with the variances $Var(x_0)=1$ and $Var(x_N)=\epsilon$. 

Note that the trusted parties must know the channel parameters to assess the security of the protocols and therefore the channel must be properly estimated. While the issue of the channel estimation was recently studied in the finite-size context \cite{Ruppert14}, in the present paper we focus on the side-channel effects and assume that the channel parameters are already known to the trusted parties. The channel estimation is still possible in the presence of the side channels because the side-channel parameters (losses and noise) can be estimated independently by the local measurements on the trusted sides. This also allows us to consider the protocols based on the preparation and measurement of a single quadrature (e.g., $x$), while the channel estimation would require additional modulation and measurement in the complementary one. Moreover, since the methods of the side-channel compensation suggested below do not change the data ensemble size (defined by the signaling and detection rate), the finite-size effects \cite{Ruppert14,finsize} would not qualitatively change the results of the paper.

Two types of side channels are considered in our study as shown in Fig. \ref{scheme}(a). The first one (further also referred to as the {\em type-A} side channel) is the sender-side side-channel leakage, when a vacuum mode is coupled to the signal and only the output of the coupling $C_A$ is accessible by an eavesdropper. An eavesdropper Eve has no control of the side-channel input and of the strength of the coupling, thus the input state of such a side channel is initially vacuum. Eve however receives the side-channel output similarly to non-invasive passive attacks in classical cryptography \cite{ClassCrypto}. The sender (Alice), on the contrary, has full control of the input of such a side channel before the coupling $C_A$. The second type of side channel (further also referred to as the {\em type-B} side channel) is untrusted noise addition in the receiver station. In this case an untrusted noise with variance $V_N$ is supposedly prepared by Eve and coupled to the signal prior to detection with the output of the coupling being inaccessible to the eavesdropper. The eavesdropper is not able to change the coupling strength. On the contrary, the receiving side (Bob) is able to control (e.g., measure) the output of the coupling $C_B$. In both the cases we assume that the trusted parties are not able to directly remove the side channels or change the coupling strengths ($C_A$ and $C_B$, respectively).

These are the two main types of possible semitrusted side channels, while the completely trusted noise is covered by  previous research \cite{Filip08,Usenko10,Cerf,Usenko2016} and completely untrusted noise can be attributed to the channel. Moreover, the noise infusion on the sender side (symmetrical to the type-B side channel on the receiver side that is considered in the present paper) is equivalent to the additional noise in the untrusted channel. At the same time the side-channel loss on the receiver side (symmetrical to the type-A side-channel loss on the sender side that is also considered here) is equivalent to the additional loss in the untrusted channel. Thus, our analysis covers the main possible semitrusted side channels based on the two-mode interaction in the prepare-and-measure CV QKD. 

We do not consider any specific physical realization of the side channels applying our analysis to the general case of semitrusted side channels based on the linear-optical mode interaction. However, the side channels can be expected in any real implementation of CV QKD in either fiber \cite{Madsen12, Jouguet2013} or free-space channels \cite{Usenko12b, Peuntinger14}, where coherent and squeezed states were successfully transferred. The sender-side leakage can take place in particular in the case of imperfect modulation, e.g., when the signal is mixed with a temporal, spectral, polarization, or spatial mode, which then leaves the sender station. The receiver-side noise infusion may, e.g., be caused by imperfect light collection from a free-space channel with a background radiation.

Linear optical crosstalk is well studied in the classical optical communications where it is present in the multiplexed channels and receivers \cite{Ho}, but was also reported 
in the quantum communications \cite{Peters}. Linear coupling represented by the beam splitter transformation is generally used to model the interaction of a quantum-optical system with the environment \cite{Haroche}. Therefore, in our work we use the typical linear optical interaction and model the mode coupling between the signal and the side channels as the beam splitters [see Fig. \ref{scheme}(b)]. The type-A side channel is modeled as coupling to a vacuum mode on a beam splitter with transmittance $\eta_A$. On the other side we model the type-B side channel as coupling to a thermal noise mode with variance $V_N$ on a beam splitter with transmittance $\eta_B$. In the case of the sender-side leakage (type-A) side channel the quadrature that enters the quantum channel is then changed to $x'_A=x_A\sqrt{\eta_A}+x_{SCA}\sqrt{1-\eta_A}$, where $x_{SCA}$ is the quadrature value of the vacuum state on the input of the beam splitter, $Var(x_{SCA})=1$. In the case of the noise-infusing (type-B) side channel the output of the quantum channel is changed as $x'_B=x_B\sqrt{\eta_B}+x_{SCB}\sqrt{1-\eta_B}$, where $x_{SCB}$ is the input noise of the type-B side channel with $Var(x_{SCB})=V_N$.

In the analysis of the negative impact of the side channels on CV QKD and the methods to compensate for such impact we mainly study the security against collective attacks, which in the asymptotic limit were shown to be no less effective than the most general coherent attacks \cite{cohatt}. In this case Eve performs the optimal collective measurement on the accessible modes after the process of bases reconciliation is completed, implying that Eve attaches a separate uncorrelated probe to each transmitted state and keeps probes in a quantum memory until she can gather additional information. To obtain simple insight into the conditions for insecurity of the protocols we also study the security against individual attacks, in which case Eve is limited by the individual measurement on the accessible modes. This weaker security analysis allows us to analytically derive the regions of insecurity of the protocols in the presence of side channels since insecurity against individual attacks implies insecurity against the more effective collective attacks. 

Following the generalization of the Czisz\'ar-K\"orner theorem \cite{CKt} on the quantum measurements performed by Devetak and Winter \cite{DW}, the protocol is secure if the mutual information between the trusted parties exceeds the information available to Eve on the data on the trusted receiver side (in the case of reverse reconciliation, which is stable against channel loss \cite{Grosshans2003a}). The security is then described by the positivity of the lower bound on the key rate which in the case of collective attacks reads:
\begin{equation}
K=\beta I_{AB}-\chi_{BE}\:,
\label{Security2}
\end{equation}
where $\beta \in (0,1)$ is the postprocessing efficiency and $\chi_{BE}$ is Holevo bound that determines Eve's achievable information limit in the case of collective attacks \cite{coll1,coll2}. The efficiency $\beta$ depends on the effectiveness of the data postprocessing algorithms that are being used in the secure key distillation given the particularly low signal-to-noise ratio. We set $\beta$ as an independent parameter and do not consider any particular postprocessing algorithm. In the following analysis we therefore fix the reconciliation efficiency as $\beta=0.95$, which is realistic taking into account the recent progress in the error-correcting algorithms for the Gaussian-distributed data \cite{errorcorr}.

The Holevo bound can be expressed as $\chi_{BE}=S(\gamma_E)-S(\gamma_{E|B})$ through the von Neumann entropy $S(\gamma_E)$ of the generally multimode state (including the side channels), which is available to Eve for the collective measurement described by the respective covariance matrix $\gamma_E$, and the von Neumann entropy $S(\gamma_{E|B})$ of the state available to Eve conditioned on the measurement results of the remote trusted party Bob \cite{Lod} and described by the covariance matrix $\gamma_{E|B}$. Covariance matrices are the matrices of the second moments of quadratures of the form $\gamma_{ij}=\langle r_ir_j \rangle - \langle r_i \rangle\langle r_j \rangle$, where $r_i=(x_i,p_i)^T$ is the quadrature vector of an $i$-th mode. Along with the first moments the covariance matrices explicitly describe the Gaussian states and are sufficient for the security analysis of the Gaussian protocols \cite{coll1,coll2} due to the extremality of the Gaussian states \cite{ExtremeG}. We analyze the security against the collective attacks using the most general purification method \cite{Lod}, where the equivalent entanglement-based representation of the protocols is used and all the state imperfections corresponding to the side channels and the main channel are attributed to Eve.

In the case of individual attacks the upper bound on the information available to an eavesdropper is given by the Shannon mutual information $I_{BE}$ instead of the Holevo bound and the lower bound on the key rate (in the optimistic case of perfect postprocessing efficiency) reads $K_{ind}=I_{AB}-I_{BE}$. Details of calculations for security analysis in the cases of both individual and collective attacks are given in the Appendix, while here we present the main expressions and results.  

In the next section we study the negative impact of the side channels on CV QKD.

\section{Negative effect of side channels.}
\label{prob}
\subsection{Side-channel loss on the trusted sender side} 
Let us first consider the type-A side channel. We start by analyzing the region of insecurity of the protocol with respect to the individual attacks and without the untrusted channel noise. The mutual information in this case reads (see the Appendix for details)
%
%
\begin{equation}
I_{AB}=\frac{1}{2}\log_{2}\frac{1}{1-\frac{\eta_A\eta V_M}{\eta_A\eta(V-1)+1}}\label{typeAmi},
\end{equation}
while the information available to Eve reads
\begin{equation}
I_{BE}=\frac{1}{2}\log_{2}\frac{[\eta_A \eta (V-1)+1][V-\eta_A \eta (V-1)]}{V}\label{typeAmiE}
\end{equation}
and is independent of the signal states (squeezed or coherent). As can be seen the side channel decreases the mutual information between the trusted parties and increases Eve's information, therefore limiting the key rate already for individual attacks with pure channel losses.

In the optimal (given perfect postprocessing $\beta=1$) limit of infinite squeezing and modulation $\left(V\rightarrow\infty\right)$ upon pure channel loss ($\epsilon=0$) the key rate for the standard Gaussian CV QKD protocol can be written as
\begin{equation}
K_{V\rightarrow\infty}=\lambda\log_{2}\frac{1}{1-\eta_A \eta},
\end{equation}
where $\lambda=1$ for the squeezed-state protocol and $\lambda=1/2$ for the coherent-state one.
The channel transmittance $\eta$ is therefore effectively decreased by the side-channel coupling $\eta_A$. Thus the presence of the type-A side channel does not break the security, i.e., the key rate remains positive for any nonzero value of $\eta_A$, as one would expect, because the channel remains purely lossy. 
%
\begin{figure}
\begin{tabular}{c}
	\includegraphics[scale=0.7]{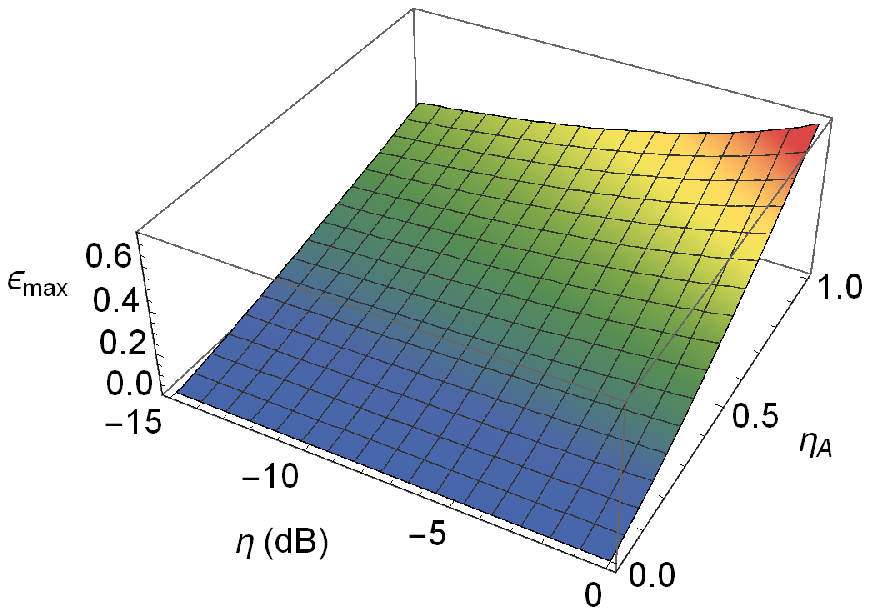} \\
	\includegraphics[scale=0.7]{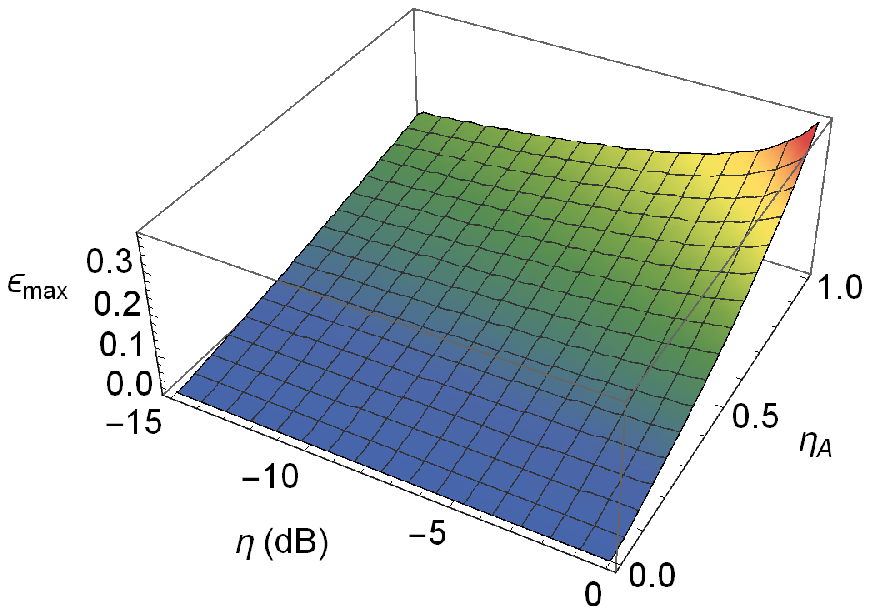} \\
\end{tabular}
	\caption{Maximum tolerable excess noise dependence on the channel losses (on a dB scale) and the type-A side-channel coupling ratio $\eta_A$ for the standard squeezed-state (top plot) and coherent-state (bottom plot) protocols. Here $\beta=1$and $V=10^3$.}
	\label{Emax1}
\end{figure}

If the channel noise is present, then the side-channel loss increases the sensitivity of the protocol to the channel noise already in the case of the individual attacks. In the limit of strong modulation $V \to \infty$ and strong channel loss $\eta \ll 1$, the maximum tolerable channel noise is $\epsilon_{max}=\eta_A/2$ for the standard coherent-state protocol and $\epsilon_{max}=\eta_A$ for the standard squeezed-state protocol with arbitrarily strong squeezing.

In the case of collective attacks (see the Appendix for details) the side-channel leakage on the trusted sender side also lowers the key rate and substantially reduces the tolerance to the channel excess noise, which is clearly visible in Fig. \ref{Emax1}, where the maximum tolerable channel excess noise $\epsilon_{max}$ (in shot-noise units, being the variance of vacuum fluctuations) is plotted versus channel transmittance and side-channel loss for the standard CV QKD protocols with strong modulation.
\subsection{Noise infusion on the trusted receiver side} 
The performance of the protocols is different in the case of the type-B side channel. In this case the presence of additional noise $V_N$ coupled to a signal can lead to the security break already for the purely attenuating channel (i.e., when $\epsilon=0$). The mutual information between the trusted parties in this case is reduced by the noise $V_N$ and reads
%
%
%
%
\begin{equation}
I_{AB}=\frac{1}{2}\log_2{
\frac{1}{1-\frac{\eta\eta_B V_M}{\eta_B(\eta V+1-\eta)+(1-\eta_B)V_N}}
}.\label{typeBmi}
\end{equation}
The security break can be observed already in the case of individual attacks upon pure channel loss. Eve's upper bound on the leaking information depends only on the overall variance $V$ and reads
\begin{equation}
I_{BE}=\frac{1}{2}\log_2{\frac{\eta_B(\eta V+1-\eta)+(1-\eta_B)V_N}{\frac{\eta_BV}{\eta+(1-\eta)V}+\frac{1-\eta_B}{V_N}}}.\label{typeBmiE}
\end{equation}
In the limit of strong modulation $V \to \infty$ and strong channel loss $\eta \ll 1$ the bound on the side-channel noise for either the squeezed- or coherent-state standard CV QKD protocol reads
\begin{equation}
V_N^{max}\big |^{V\to\infty}_{\eta\ll 1}=\frac{1}{1-\eta_B}.
\end{equation}
%
%
\begin{figure}
\begin{tabular}{c}
	\includegraphics[scale=0.7]{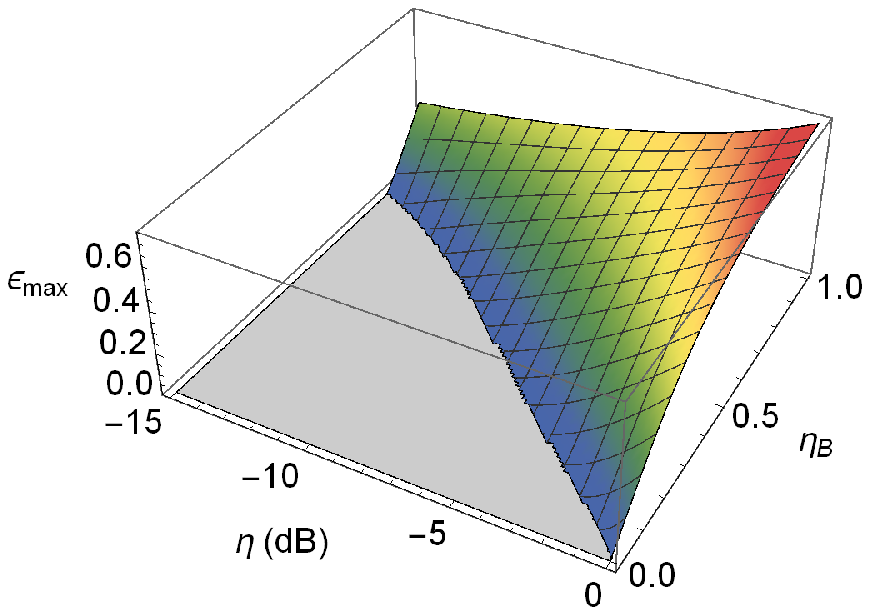} \\
	\includegraphics[scale=0.7]{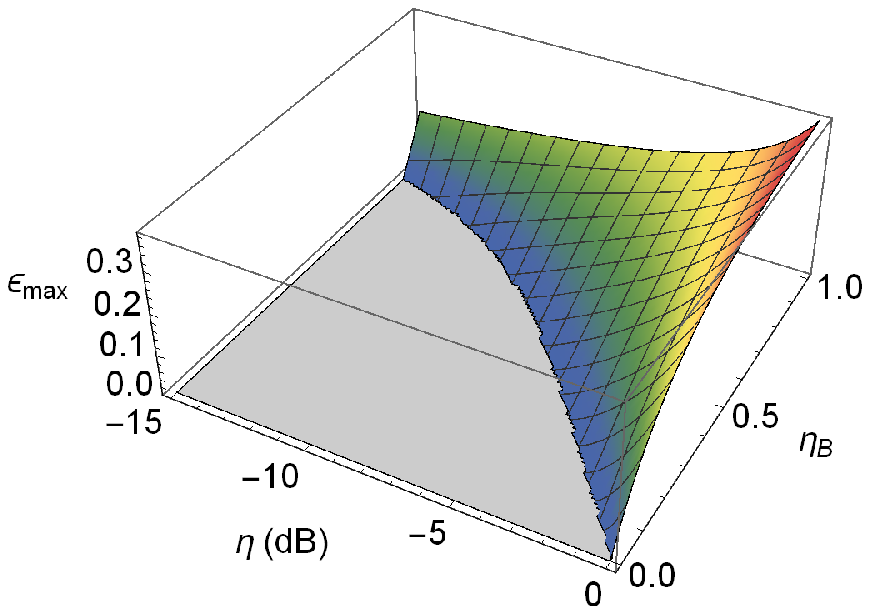} \\
\end{tabular}
	\caption{Maximum tolerable excess noise dependence on the channel losses (on a dB scale) and the type-B side-channel coupling ratio $\eta_B$ for the standard squeezed-state (top plot) and coherent-state (bottom plot) protocols. Here $\beta=1$, $V=10^3$, and the side-channel noise variance $V_N=1.05$.}
	\label{EmaxBN}
\end{figure}

In the more general case of collective attacks the side-channel noise $V_N$ not only undermines the tolerance of the protocol to the channel noise $\epsilon$, but also leads to the security break contrary to the type-A side-channel leakage. This can be seen from the profiles of the maximum tolerable channel noise in the case of $V_N=1.05$, i.e., when the input of the type-B side channel only slightly exceeds the shot-noise variance as shown in Fig. \ref{EmaxBN}. Note that the squeezed-state protocol appears to be more stable against the side-channel noise infusion (its security region is larger in terms of the tolerable channel loss and side-channel coupling at the given $V_N$). Thus, the presence of the side-channel leakage or noise infusion makes the protocol more sensitive to the channel noise and can even break the security for the purely attenuating channel. In the next section we suggest the methods to compensate for negative effects by manipulations at the trusted sides and without affecting the untrusted quantum channel.

\section{Decoupling of side channels.}
\label{solve}
\subsection{Side-channel loss on the trusted sender side}
\begin{figure}
	\centering
		\includegraphics[scale=0.55]{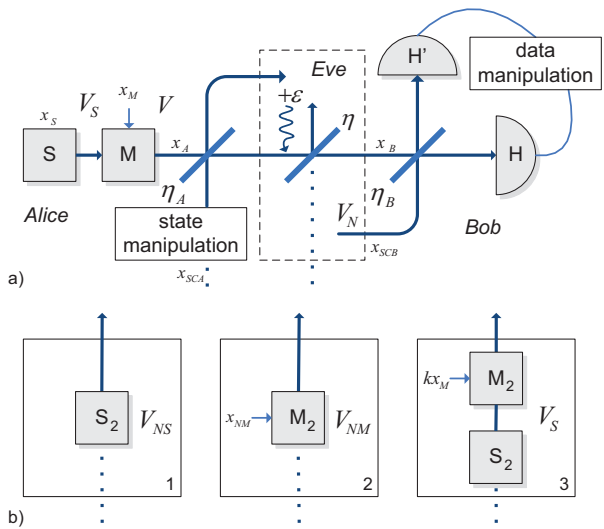}
	\caption{(a) Methods aimed at compensating for the negative impact of the side channels: state manipulation on the input of the type-A sender-side lossy side channel and monitoring of the output of the type-B receiver-side noise-infusing side channel using the monitoring homodyne detector $H'$ and subsequent data manipulation involving also the measurement results from the main homodyne detector $H$. (b) Types of state manipulation on the input of the type-A side channel: 1 (left), noise infusion using the source $S_2$ producing a thermal state with variance $V_{NS}$; 2 (middle), controllable uncorrelated modulation on the side-channel input using the modulator $M_2$; and 3 (right), controllable correlated modulation with displacement $kx_M$ proportional to the modulation of the main signal using the modulator $M_2$. In the case of the squeezed-state protocol to achieve complete decoupling of the side channel the side-channel input should be replaced by the squeezed state with variances $V_S$ and $1/V_S$ using the source $S_2$ prior to the modulator $M_2$. In the case of the coherent-state protocol such preparation is not needed and the source $S_2$ needs not be used.}
	\label{solution}
\end{figure}

We suggest that the trusted sender (Alice) should look for the input of the type-A side channel in the case it cannot be removed completely and then apply state manipulation on the side-channel input [see Fig. \ref{solution}(a)]. Three options can be considered depending on the accessibility of the side channel and technical ability of Alice. 

\begin{figure}[t]
\begin{tabular}{ll}
	\includegraphics[scale=0.45]{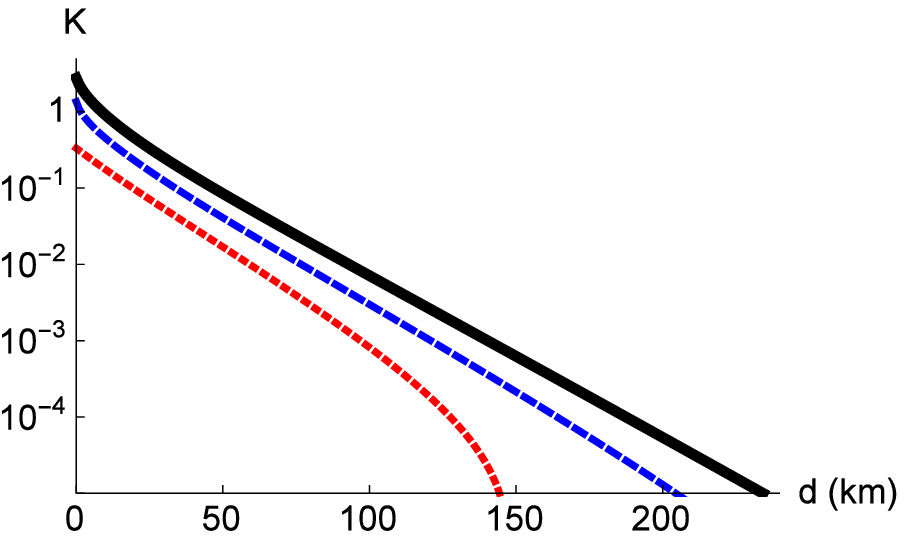} 
	\includegraphics[scale=0.45]{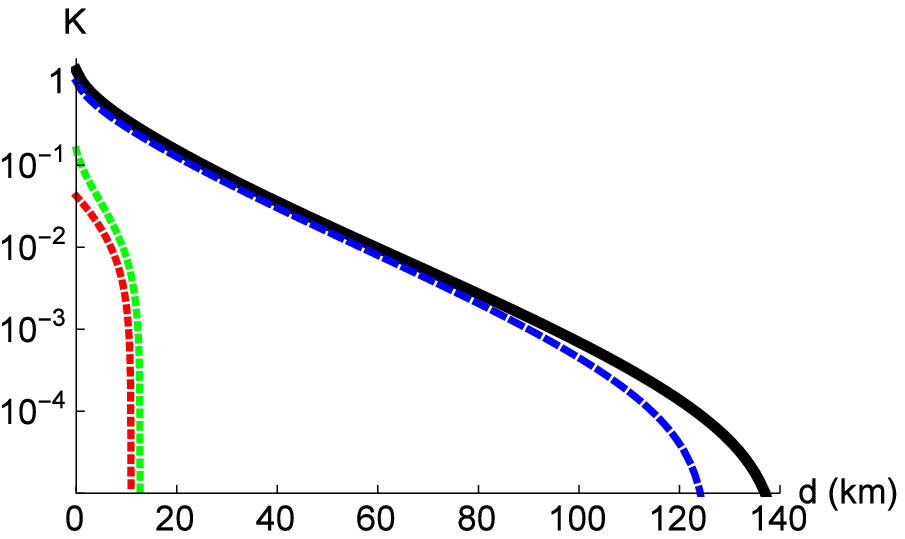}
\end{tabular}
	\caption{Key rate secure against collective attacks versus distance in a standard telecom fiber (with an attenuation of $-0.2 dB/km$) for the squeezed-state protocol with $V_S=0.1$ (left) and the coherent-state protocol (right) in the presence of the type-A side channel $\eta_A=0.4$ and no compensating methods (red dotted lines), with optimized unknown noise on the input of the channel (green upper dotted line for the coherent-state protocol), with optimized uncorrelated modulation on the input of the side channel (blue dashed lines), with optimized correlated modulation on the input of the side channel and no additional source $S_2$ (coincides with the blue dashed line for the squeezed-state protocol), and in the perfect case in the absence of the side channel, i.e., $\eta_A=1$ (solid black lines). The latter curve overlaps with the ones for the optimized correlated modulation for the coherent-state protocol; for the optimized correlated modulation and squeezing of the side-channel input; and for the optimized uncorrelated modulation and squeezing on the input of the side channel for the squeezed-state protocol. Here $\beta=0.95$, $\epsilon=5\%$, and the modulation variance $V_M$ is optimized for the given parameters.}
	\label{Distance}
\end{figure}

First, Alice can infuse Gaussian thermal noise to the side channel by replacing the vacuum input of the side channel with the source of noise with variance $V_{NS}$ [see Fig. \ref{solution}(b), left]. The efficiency of such method is however very limited. Indeed, such noise reduces the mutual information
\begin{equation}
I_{AB}=\frac{1}{2}\log_2{\frac{1}{1-\frac{\eta_A\eta V_M}{\eta[\eta_A V+(1-\eta_A)V_{NS}]+1-\eta}}}\label{typeAmi1}.
\end{equation}
However, it also, to some extent, decreases the Holevo quantity due to a partial decoupling of the side channel from the main channel, but at the same time acts as a preparation noise \cite{Usenko10}. Thus, the addition of such unknown noise is of limited helpfulness, when the main channel has low loss, i.e., is short distance. Moreover, for the squeezed-state protocol, where the Holevo bound is effectively minimized by squeezing \cite{Usenko11}, the reduction of the mutual information due to the presence of additional noise appears to be more harmful, so mostly the unknown noise on the input of the side channel has either no or a very limited positive effect. This can be seen from the graphs in Fig. \ref{Distance}, where the key rate is plotted versus distance $d=-50\log_{10}{\eta}$ in a standard telecom fiber with attenuation of $-0.2 dB/km$ (here and in the following we plot the key rate in bits per measurement). The improvement for the coherent-state protocol is small but visible [upper (green) dotted line compared to the lower (red) dotted one], while the improvement for the squeezed-state protocol is negligible (the corresponding curve overlaps with the one with no manipulation on the side-channel input performed, given as the dotted red line).

Second, Alice can use the additional modulator $M_2$ on her side to control the input of the side channel. Alice's modulation therefore shifts the quadrature of the side-channel input $x_{SCA}$. Let us assume that the additional modulation (displacement) on the input of the type-A side channel is independent from the main modulation performed on the signal, but is known to Alice and contributes to her data and to the correlation with Bob [see Fig. \ref{solution}(b), center]. We can write the change of the input of the lossy side channel in terms of the $x$ quadrature (calculations for the case when the $p$ quadrature is modulated and measured will be equivalent) as $\tilde{x}_{SCA}=x_{SCA}+x_{NM}\:,$ where $x_{NM}$ is the shift, known to Alice, with variance $Var(x_{NM})=V_{NM}$. 
The mutual information between the trusted parties in this case is increased:
\begin{equation}
I_{AB}=\frac{1}{2}\log_2{\frac{1}{
1-\frac{
\eta(\sqrt{\eta_A}V_M+\sqrt{1-\eta_A}V_{NM})^2
}{(V_M+V_{NM})\big(
\eta[\eta_A(V-V_{NM})-\eta_A+\epsilon]+1\big)
}
}}\label{typeAmi2}
\end{equation}
due to increased correlations between the trusted parties. However, it simultaneously decorrelates (reduces the correlation with the main signal mode) the output of the side channel and increases the information leakage from the main channel. Therefore, such additional uncorrelated modulation on the input of the side channel $V_{NM}$ can play a positive role mainly when the side channel is strong enough (typically $\eta_A<0.8$) because otherwise the information leakage from the main channel prevails over the positive role of decoupling. Moreover, the modulation variance $V_{NM}$ must be optimized for the given setup parameters. However, such a method can significantly increase the secure distance of the protocol especially for the coherent-state protocol, as can be seen from Fig. \ref{Distance}, where the corresponding key rate is given as the blue dashed lines.

Third, we suggest the method of correlated modulation on the input of the side channel and optionally additional squeezing of the side-channel input in the case of the squeezed-state protocol. Importantly, the method uses only the classical correlation of the Gaussian quantum states; no entanglement is required. The method as we show below allows (i) complete decoupling of the modulation from the side channel (no fraction of the modulation data appears on the side-channel output) and (ii) complete decorrelation of the side-channel output from the signal mode. These effects allow one to restore the performance of the protocol and thus completely remove the negative influence of the type-A side channel. 

Indeed, Alice can apply the weighted correlated displacement on the input of the side channel [see Fig. \ref{solution}(b), right] with the factor $k$ so that the input of the side channel becomes $\tilde{x}_{SCA}=x_{SCA}+kx_M$. After the coupling between the signal and the modulated side-channel input the quadratures are $x'_A=x_S\sqrt{\eta_A}+x_{SCA}\sqrt{1-\eta_A}+x_M(\sqrt{\eta_A}+k\sqrt{1-\eta_A})$ and $\tilde{x}'_{SCA}=x_{SCA}\sqrt{\eta_A}-x_S\sqrt{1-\eta_A}+x_M(k\sqrt{\eta_A}-\sqrt{1-\eta_A})$. It is easy to see that when $k=\sqrt{(1-\eta_A)/\eta_A}\equiv k_{opt}$ the outputs of the side-channel coupling become $x'_A=x_S\sqrt{\eta_A}+x_{SCA}\sqrt{1-\eta_A}+x_M/\sqrt{\eta_A}$ and $\tilde{x}'_{SCA}=x_{SCA}\sqrt{\eta_A}-x_S\sqrt{1-\eta_A}$. Therefore, due to the destructive interference effect, the untrusted output of the side channel contains no information on the signal displacement $x_M$, i.e., the side channel is completely decoupled from the modulation. Then, in the case of the coherent-state protocol since $Var(x_{SCA})=V_S=1$ the correlation between the outputs of the side-channel in the regime of optimal correlated modulation with $k_{opt}$ vanishes, i.e., $Cov(x'_A,\tilde{x}'_{SCA})=0$ and the output of the side channel, containing already no encoded information, becomes in addition completely decorrelated from the signal mode. The eavesdropper therefore cannot profit from such the side channel. Importantly, both conditions above (decoupling the modulation and decorrelating the side channel) are required to fully eliminate the side-channel effect. In the case of the squeezed-state protocol the decorrelation is achieved upon the additional manipulation on the input of the side channel prior to the modulation so that the vacuum state is replaced by the squeezed state with variances $(V_S,1/V_S)$, equivalent to the signal state. Using this generated squeezing in addition to the optimal correlated modulation, the output of the side channel is completely decorrelated from the signal mode for the squeezed-state protocol as well. 

\begin{figure}
	\centering
		\includegraphics[scale=0.7]{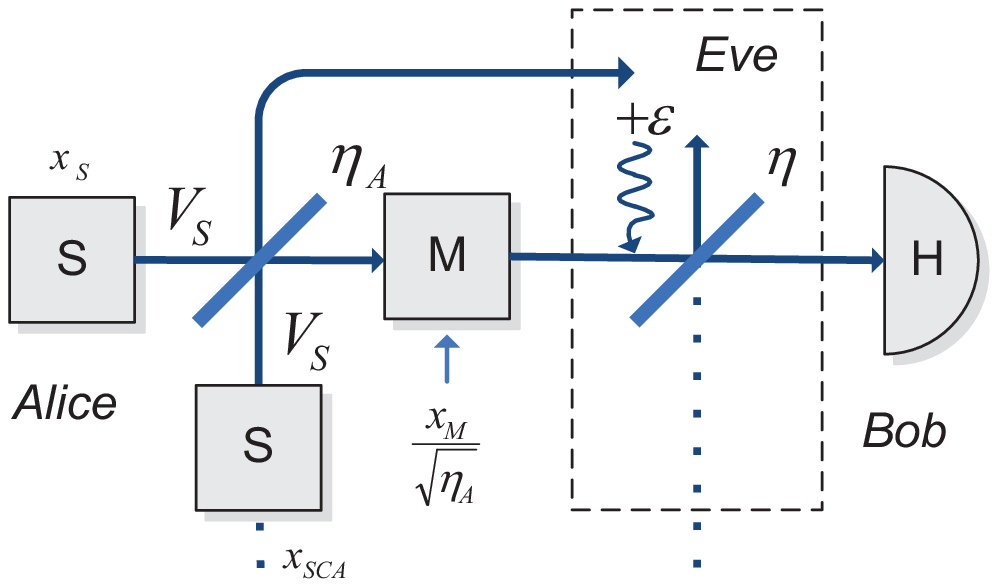}
	\caption{Equivalent scheme of the method [depicted in Fig. \ref{solution}(b), right], in the case of the optimal correlated displacement with $k=\sqrt{(1-\eta_A)/\eta_A}$ applied to the input of the side channel. The side channel is effectively moved to the signal state prior to the main modulator $M$ and the displacement on the signal is scaled as $x_M/\sqrt{\eta_A}$. The source $S_2$ should be present in the squeezed-state protocol to achieve the complete decoupling of the side channel.}
	\label{scheme-corr}
\end{figure}

Interestingly, in the regime of the optimal modulation with $k_{opt}$ the scheme becomes equivalent to the side-channel attack on the signal prior to modulation; the latter then becomes scaled by $1/\sqrt{\eta_A}$ as shown in Fig. \ref{scheme-corr}. In other words, the optimal correlated displacement with $k_{opt}$ shifts the side-channel attack from the modulated signal to the signal state before the modulation. This is in fact an additional type of side-channel attack that can also take place independently of any other side-channel attacks. It is easy to see that in the case of the coherent-state protocol such an attack yields no additional information for Eve because the correlation between the output of the side channel and the signal mode after the interaction $C_{AS_A}=\sqrt{\eta_A(1-\eta_A)}[V_S-Var(x_{SCA})]$, which is proportional to the difference of variances of the incoming modes, becomes exactly zero [similarly to the method of decoupling Eve from the main quantum channel (see \cite{Jacobsen2014})]. In the case of the squeezed signal, however, such an attack on the signal states leads to the nonzero correlation between the signal state and the side-channel output and this reduces the security of the squeezed-state protocol, which, nevertheless, remains superior to the coherent-state one in terms of key rate, distance, or tolerable excess noise. Therefore, e.g., for $V_S=0.1$ the optimally correlated displacement appears to be less effective than the uncorrelated one (Fig. \ref{Distance}). This can be overcome if Alice is able to substitute the vacuum input of the side channel by a squeezed state with the same squeezing as the signal state, i.e., $Var(x_{SCA})=V_S$ should hold. In this case the correlations between the squeezed signal states and the side-channel output upon $k_{opt}$ vanish and the type-A side channel can be fully decoupled for the squeezed-state protocol as well. For details of the calculations see the Appendix.

The correlation between Alice and Bob in the regime of optimal modulation with $k_{opt}$ changes to $V_M/\sqrt{\eta_A}$ (prior to the main channel). Thus the key rate for the same $V_M$ in the regime of complete decoupling of the type-A side channel is quantitatively different from the key rate of the protocol with the same modulation and in the absence of the side channel. However, in the regime of imperfect postprocessing, i.e., $\beta<1$, the modulation variance needs to be optimized. With this optimization performed the protocol with the complete decoupling of the type-A side channel becomes fully equivalent in terms of the maximum key rate, tolerable channel loss (or, equivalently, maximum distance), and tolerable channel excess noise to the protocol without the type-A side channel and with optimized modulation for a given $\beta$. This leads in particular to the overlap of the curves for the two protocols in Fig. \ref{Distance}, where optimized key rates for the methods of the non-correlated modulation and of the unknown noise infusion are also given for comparison. Therefore by optimal decoupling of the type-A side channel using only the correlated modulation and optionally squeezing on the input of the side channel one can {\em completely} remove its negative influence with no entanglement between the main channel and side channel being required.  %
\begin{figure}[h]
\begin{tabular}{ll}
	\includegraphics[scale=0.45]{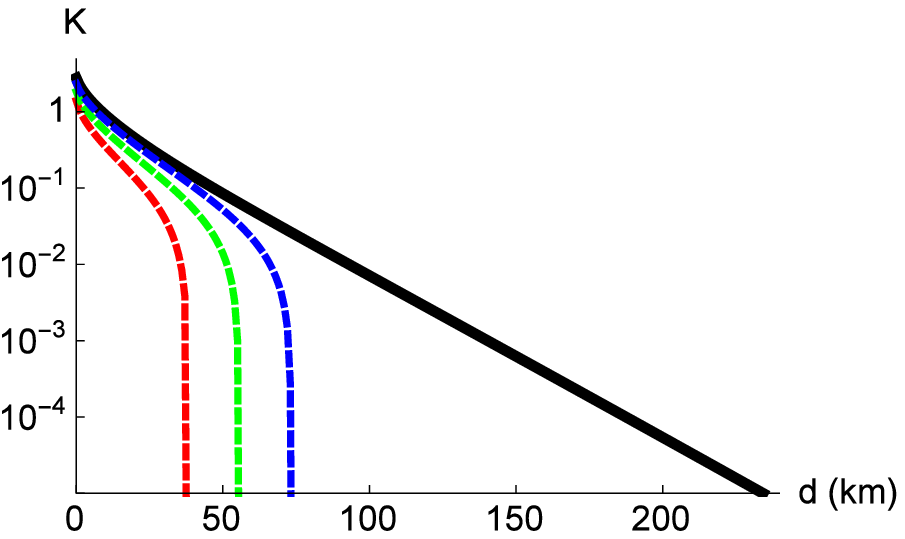} 
	\includegraphics[scale=0.45]{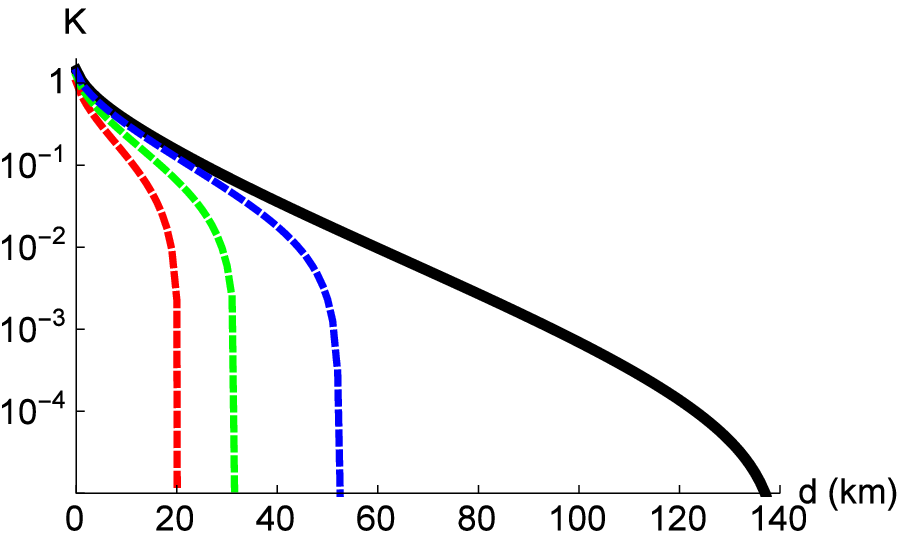} 
\end{tabular}
	\caption{Key rate secure against collective attacks versus distance in a standard telecom fiber (with an attenuation of $-0.2 dB/km$) for the squeezed-state protocol with $V_S=0.1$ (left) and the coherent-state protocol (right) in the presence of the type-B noise-infusing side channel on the receiver side without the side-channel monitoring (dashed lines) and with optimal monitoring, perfectly coinciding with the profile of the key rate without the side channel (solid black lines). The side-channel coupling is $\eta_B=0.5,0.7,0.9$ (from left to right, i.e., red, green, and blue dashed lines respectively), $\beta=0.95$, $\epsilon=5\%$, $V_N=1.05$, and the modulation variance $V_M$ is optimized for the given parameters.}
	\label{KRBS}
\end{figure} 
Note that the uncorrelated modulation can be also combined with squeezing on the input of the side channel. This combination in the case of the squeezed-state protocol greatly improves the method of uncorrelated modulation, making it (provided the modulation is optimized) almost as effective as the method of optimized correlated modulation combined with squeezing (on the plots in Fig. \ref{Distance} the corresponding line in the plotted region of parameters overlaps with the black solid line corresponding to the absence of the side channel and the difference corresponding to the limited performance of the method can only be seen for very low values of the key rate, which are irrelevant due to unavoidable finite-size effects \cite{Ruppert14,finsize}).
\subsection{Noise infusion on the trusted receiver side} 
In the case when Eve couples an additional noise to the signal prior to the detection at Bob's side, the monitoring of the coupling output, which is not accessible to Eve, can be used. Then, by applying the proper manipulation on the data from the main detector and from the monitoring detector the negative influence of the type-B side channel can also be fully compensated for.

We suggest the method of weighted subtraction of data from the main and the monitoring detector and show that the resulting measurement is free from the influence of the type-B side channel. Indeed, if the main homodyne detector $H$ (see Fig. \ref{solution}) after the noise-infusing side channel measures the quadrature $x'_B=x_B\sqrt{\eta_B}+x_{SCB}\sqrt{1-\eta_B}$, where $x_B$ is the output of the main quantum channel and $x_{SCB}$ is the noise quadrature of the type-B side channel input with $Var(x_{SCB})=V_N$, and the monitoring detector $H'$ measures the quadrature $x'_{SCB}=-x_B\sqrt{1-\eta_B}+x_{SCB}\sqrt{\eta_B}$, then the weighted difference $\Delta x=gx'_B-g'x'_{SCB}$ (and similarly for the $p$ quadrature) is free from the influence of the side channel for $g=\sqrt{\eta_B}$ and $g'=\sqrt{1-\eta_B}$. Therefore, the additional optimized monitoring on the output of the noise-infusing side channel resulting in the detection of $\Delta x=x_B$ can {\em completely} remove the negative impact of such a side channel. Note that any pair of coefficients satisfying $g/g'=\sqrt{\eta_B/(1-\eta_B)}$ fully restores the performance of the protocol leading to $\Delta x\propto x_B$, and the linear scaling of the latter observable does not affect the lower bound on the secure key rate.

The complete removal of the noise-infusing side channel is possible also with the imperfect detectors. If both the main homodyne detector $H$ and the monitoring detector $H'$ have efficiency $\eta_D$ and excess noise, which can be modeled by coupling of the signal to the noise mode with the variance $V_D$ on the coupler $\eta_D$ (this is the standard model of the imperfect homodyne detector used in the security analysis of CV QKD \cite{Lod}), then the settings $g$ and $g'$ given above also remove $x_{SCB}$ from the weighted difference $\Delta x$ and the variance then reads $Var(\Delta x)=\eta_D Var(x_B)+(1-\eta)V_D$. That is, the optimal monitoring of the side-channel output then becomes equivalent to the side-channel-free detection of the signal on the same imperfect homodyne detector (the details of the calculations are given in the Appendix), which contains only the trusted noise and thus does not lead to the security break in the reverse-reconciliation scheme \cite{Cerf}.

To calculate the security against the collective attacks we consider the scheme using the equivalent interferometric setup, when the residual side channel is coupled to the signal and then detected (see the Appendix for details). This leads to the appropriate transformations of the variances and correlations.  The results of calculations are given in Fig. \ref{KRBS} without the side-channel monitoring and with optimal monitoring of the residual side channel. 

It is evident that the optimal side-channel monitoring restores the performance of the protocol providing exactly the same key rate as in the absence of the side channel. The noise-infusing side channel can therefore be {\em completely} compensated for. Simultaneously, the Gaussian entanglement between the trusted parties is fully restored even if it was previously broken by the effect of noise infused in the side channel. Experimental aspects of noise cancellation by the measurement have been studied in \cite{Lassen2013}, which demonstrates the feasibility of such a method for CV QKD. Note that the result reported here is obtained under different conditions than the previous analysis of the multimode channels \cite{Filip2005,Usenko2014}, where the auxiliary channels received by Bob contain information encoded by Alice, i.e., are parallel to the main quantum channel. 
\begin{figure}
	\centering
		\includegraphics[scale=0.55]{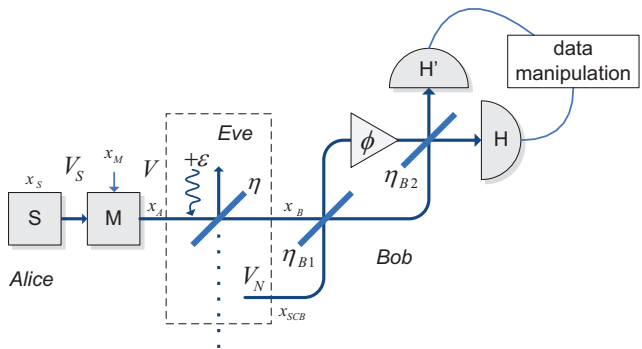}
	\caption{Scheme of the CV QKD with the generalized interferometric coupling (parametrized by the transmittance values $\eta_{B1}$ and $\eta_{B2}$ of the couplers and by the phase shift $\phi$) to the noisy side channel with untrusted input of variance $V_N$ on the receiving side and the monitoring of the side-channel output followed by the data manipulation.}
	\label{scheme-B-adv}
\end{figure}

\begin{figure}
	\centering
		\includegraphics[scale=0.8]{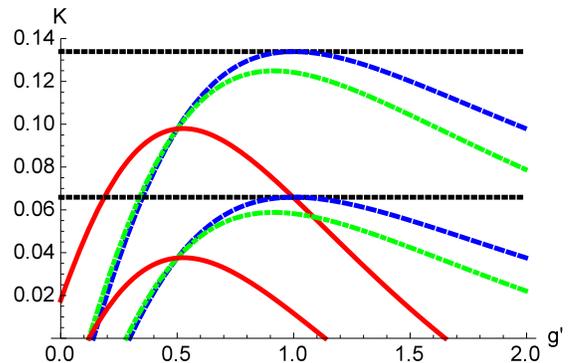}
	\caption{Key rate secure against collective attacks versus weighting of the data manipulation in the monitoring of the type-B side channel with the generalized interferometric interaction for the squeezed-state protocol with $V_S=0.1$ (upper lines) and the coherent-state protocol (lower lines) in the absence of the type-B side channel (black horizontal dotted lines), with no phase shift $\phi=0$ (dashed blue lines), with $\phi=0.5$ (green dot-dashed lines), and $\phi=1.5$ (red solid lines). The parameters of the interaction in the presence of the side channel are $\eta_{B1}=0.9$ and $\eta_{B2}=0.8$, the modulation variance $V_M=10$, the channel transmittance is $\eta=0.1$, and the protocols implementation is otherwise perfect.}
	\label{KR-B-adv}
\end{figure}

We also consider the side-channel noise infusion based on the generalized interferometric interaction modeled by two couplers with different transmittance values $\eta_{B1}$ and $\eta_{B2}$ and a phase shift $\phi$ in one of the arms between the couplers as shown in Fig. \ref{scheme-B-adv}. In this case the monitoring of the side-channel output suggested above can fully restore the performance of the protocol only when the phase shift is absent (i.e., $\phi=0$, see the Appendix for the details) and the optimal coefficients of the data manipulation read $g=1$ and $g'=\big(\sqrt{(1-\eta_{B2})\eta_{B2}}+\sqrt{(1-\eta_{B1})\eta_{B1}}\big)/(1-\eta_{B1}-\eta_{B2})$. The setting can be obtained by maximizing the mutual information between the trusted parties and therefore does not require the estimation of $\eta_{B1}$ and $\eta_{B2}$ independently. However, when the nonzero phase shift is present and the output of the interferometric coupling contains combinations of $x$ and $p$ quadratures of the signal and the noise input, simple side-channel monitoring by the homodyne detection in the $x$ quadrature and the linear data manipulation are not sufficient to completely restore the performance of the protocol. It can can be used to partly compensate for the negative influence of the type-B side channel, as shown in Fig. \ref{KR-B-adv}, where the key rate is plotted for the coherent- and squeezed-state protocols with respect to the weighting $g'$ (assuming $g=1$), which can maximize the mutual information and, respectively, the key rate.
The optimal data manipulation setting in the general case becomes the lengthy function on the parameters of the protocol, including the values of the coupling and the phase shift as well as the signal and modulation variances and the parameters of the channel. In order to improve the decoupling of the side channel an optimal additional phase shift can be applied prior to the control detection (so that the mutual information is maximized) or a more general strategy based on the heterodyne detection and subsequent data manipulation could be used and optimized similarly to the elimination of the cross-talk in the channel \cite{Filip2005}. 

\section{Conclusions}\label{concl}We have studied the effect of the side-channel leakage and noise infusion on the trusted sides of the continuous-variable quantum key distribution protocols. The negative effect of the side-channel leakage on the trusted sender side leads to the degradation of the key rate and to the increased sensitivity of the protocol to the channel noise. At the same time, the side-channel noise infusion on the trusted receiver side can completely break the security of the protocols even upon pure channel loss. We suggested and examined the method of additional modulation applied to the side-channel input being under the control of a trusted sending party. We show that if the additional modulation is properly correlated with the main modulation on the signal and squeezing applied on the side-channel input in the case of the squeezed-state protocol, the negative impact of the lossy side channel can be completely removed. Alternatively, we show the possibility to compensate for the negative impact of the noisy side channel on the receiver side by introducing the monitoring of the output of the side channel. Since both methods work independently by completely removing the side channels, they can be combined in a single protocol. Moreover, since the optimal settings for the methods are independent of the channel parameters, the methods can be applied by the trusted parties using only the parameters of their local trusted stations and do not themselves rely on the channel estimation. Our result therefore describes the effective and feasible methods of compensating for the quantum side channels in a continuous-variable quantum key distribution between the trusted parties, which do not require entanglement or non-Gaussian operations.

\medskip
\noindent {\bf Acknowledgments} The research leading to these results has received funding from the EU FP7 under Grant Agreement No. 308803 (Project BRISQ2), cofinanced by M\v SMT \v CR (Grant No. 7E13032). I.D. acknowledges Palack\'y University Project No. \text{IGA\_PrF\_2014008}. I.D. and V.C.U. acknowledge the Project No. P205/12/0694 of the Czech Science Foundation.


\appendix*
\section{Security analysis is detail}
\label{appendix}
Here we provide detailed calculations for the security analysis of the above-described Gaussian continuous-variable quantum key distribution protocols with side channels. 
\subsection{Scheme and parametrization} 
The scheme of the protocols is given in Fig. \ref{scheme}. As mentioned, the channel is parametrized by transmittance (loss) $\eta$ and excess noise $\epsilon$, while side channels are parametrized by coupling $\eta_A$ (for the sender-side type-A lossy side channel) and by coupling $\eta_B$ and excess noise $V_N$ (for the receiver-side type-B noise-infusing side channel). The protocols in the prepare-and-measure (PM) setting are based on the preparation of a signal state (coherent or squeezed) characterized by the quadrature values $x_S$ and $p_S$ which are Gaussian distributed around zero with variances $Var(x_S)=V_S$, $Var(p_S)=1/V_S$, where $V_S \leq 1$ is generally the squeezed variance which in the case of coherent states is saturated by $V_S=1$. Here and in the following, with no loss of generality, we assume that the states are squeezed and measured in the $x$ quadrature. The results for the $p$ quadrature squeezing and measurement are obtained by replacing $x \to p$ and vice versa. The signal is modulated by applying the displacement $x_M$ and $p_M$ randomly chosen from a Gaussian distribution centered around zero with variance $Var(x_M)=Var(p_M)=V_M$ so that the resulting quadrature becomes $x_A=x_S+x_M$. Here and in the following the equivalent expressions apply to the $p$ quadrature since the main quantum channel and the side channels are assumed to be phase insensitive (which is valid for typical optical channels such as optical fiber or free-space links). Now if the channel is present then quadrature values after the channel are given by $x_B=(x_A+x_N)\sqrt{\eta}+x_0\sqrt{1-\eta}$, where $x_0$ is the quadrature value of the vacuum state coupled to the signal to describe the loss, $Var(x_0)=1$ and $x_N$ is the quadrature value of the excess noise, $Var(x_N)=\epsilon$. If the side-channel loss is present at the sender side (type-A side channel), then the signal is coupled to the vacuum input of the side channel, which is modeled by a beam splitter with transmittance $\eta_A$, which is the side-channel loss. As it was mentioned, the quadrature that enters the quantum channel is then changed to $x'_A=x_A\sqrt{\eta_A}+x_{SCA}(\sqrt{1-\eta_A})$, where $x_{SCA}$ is the quadrature value of the vacuum state on the input of the beam splitter, $Var(x_{SCA})=1$. If the noise-infusing side channel is present, then, as mentioned in the main text, the output of the quantum channel is further modified to $x'_B=x_B\sqrt{\eta_B}+x_{SCB}\sqrt{1-\eta_B}$, where $x_{SCB}$ is the input noise of the type-B side channel with $Var(x_{SCB})=V_N$. Knowing the transformation of the quadrature values we can obtain the variances and correlations between the quadratures and derive the covariance matrices describing the states shared between the trusted parties Alice and Bob and available to an eavesdropper Eve, which are then used in the security analysis below.
\subsection{Secure key rate} 
As mentioned, we estimate the security of the protocols in the presence of the side channels and upon additional manipulations aimed to remove the side channels, as the value and positivity of the lower bound on the key rate, which in the case of collective attacks (when Eve is able to collectively measure her probe states after interaction with the signal) and reverse reconciliation \cite{Grosshans2003a} reads $K=\beta I_{AB}-\chi_{BE}$, where $\beta \in (0,1)$ is the postprocessing efficiency that takes into account the amount of data that trusted parties lose due to imperfections of the error correction algorithms, $I_{AB}$ is the mutual information between the trusted parties, and $\chi_{BE}$ is the Holevo bound, giving the upper bound on the information that is available to Eve. In the case of individual attacks, when Eve is limited by the individual measurement on her probe states, the Holevo bound is replaced by the classical Shannon information between Eve and Bob $I_{BE}$.
\subsection{Mutual information and individual attacks}
In order to calculate the classical (Shannon) mutual information we use the expression for Shannon entropies in the case of Gaussian continuous distributions  \cite{Shannon}
\begin{equation}
I_{XY}=\frac{1}{2}\log_2{\frac{V_{X}}{V_{X|Y}}},\label{ixy}
\end{equation}
where $X$ and $Y$ are two zero-mean Gaussian random variables with variances $V_X \equiv \langle X^2 \rangle$ and $V_Y \equiv \langle Y^2 \rangle$, respectively, and $V_{X|Y}=V_X-C_{XY}^2/V_Y$ is the conditional variance with $C_{XY} \equiv \langle XY \rangle$ the correlation (covariance) between $X$ and $Y$. Note that (\ref{ixy}) is symmetrical with respect to $X$ and $Y$. In the case of the Gaussian protocols considered in the paper, the variables are the quadratures displacements being introduced by modulation and the quadrature values measured on the remote side of the channel and by a potential eavesdropper are all Gaussian distributed. This allows us to calculate the mutual information $I_{AB}$ and upper bound the information leakage $I_{BE}$ in the case of individual attacks.

The calculation of the mutual information is straightforward. Following the expression (\ref{ixy}), we can derive the mutual information between Alice and Bob as $I_{AB}=\frac{1}{2}\log_2{(V_A/V_{A|B})}$,
where $V_A$ is the variance of the data imposed by Alice by displacement (typically equivalent to $V_M$), while conditional variance $V_{A|B}=V_A-C_{AB}^2/V_B$ involves correlation $C_{AB}=Cov(x_M,x_B)$, i.e., the covariance between the data kept by Alice and the data measured by Bob, and the variance $V_B=Var(x_B)$ of Bob's measurement results (which is $x'_B$ if the type-B side channel is present).

The calculation of Eve's information, $I_{BE}$, in the case of individual attacks, is similar. It requires knowing the variances of the modes that are available to Eve for the individual measurements and correlations with the measurement results on the side of Bob; these will be derived in the particular cases below. 
\subsection{Collective attacks} 
In the case of collective attacks, as mentioned, the information, which is available to Eve, is bounded by the Holevo quantity, which is the capacity of a bosonic channel between Eve and Bob. It is calculated as $\chi_{BE}=S(\gamma_{E})-S(\gamma_{E|B})$, the difference of the von Neumann (quantum) entropies $S(\gamma_{E})$ of the state of the modes, which are available to Eve for a collective measurement described by the covariance matrix $\gamma_E$, and $S(\gamma_{E|B})$ of the same state conditioned on the measurement results of Bob.  

In the general case, when the excess noise is present in the channel and/or in the type-B side channel, we use the purification method \cite{Lod}, i.e., we assume that an eavesdropper Eve can purify the state shared by the trusted parties, so $S(\gamma_{E})=S(\gamma_{AB})$, where AB is generally a multimode initially pure state shared between the trusted parties in which all the impurity is assumed to be caused by Eve's collective attack. After Bob's projective measurement of one of the quadratures similar equivalence holds for the conditioned states: $S(\gamma_{E|B})=S(\gamma_{A|B})$. Thus the Holevo bound in Eq. (\ref{Security2}) is expressed as $\chi_{BE}=S(\gamma_{AB})-S(\gamma_{A|B})$. 

The entropy $S(\gamma_{AB})$ is determined from the symplectic eigenvalues $\lambda_{i}$ of the $n$-mode covariance matrix $\gamma_{AB}$ as 
\begin{equation}
S=\sum_{i=1}^n G\left[\frac{\lambda_{i}-1}{2}\right]\:,\label{eq:von_Neumann}
\end{equation}
where $G$ is the bosonic entropic function \cite{GaussRev}
\begin{equation}
G\left(x\right)=\left(x+1\right)\log_{2}\left(x+1\right)-x\log_{2}\left(x\right)\label{eq:BosonicEF}.
\end{equation}
The subtrahend in the expression for the Holevo bound is the entropy similarly determined by the symplectic eigenvalues of the respective conditional covariance matrix $\gamma_{A|B}$ after Bob's projective measurement (with no loss of generality, we assume measurement of the $x$ quadrature):
\begin{equation}
\gamma_{A|B}=\gamma_{A}-\sigma_{B|A}\left[X\gamma_{B}X\right]^{MP}\sigma_{B|A}^T\:, \label{Conditional_CovM}
\end{equation}
where $\sigma_{B|A}$ is the correlation matrix between mode B and the rest of the trusted modes, $X=\text{Diag}(1,0,0,0)$, where $\text{Diag}$ denotes a diagonal matrix, and $MP$ is the Moore-Penrose pseudoinverse of the matrix. 

The purification \cite{Lod} is typically based on introducing the entangled [also referred to as Einstein-Podolsky-Rosen (EPR)] sources, which are the two-mode vacuum states described by the covariance matrices of the form
\begin{equation}
\gamma_{EPR}=
\left(
\begin{array}{cc}
V\mathbb{I} & \sqrt{V^2-1} \sigma _z \\
\sqrt{V^2-1}\sigma _z & V\mathbb{I} \\
\end{array}
\right),
\end{equation}
where $V$ is the variance of each of the two modes, $\mathbb{I}$ is the 2x2 unity matrix and $\sigma_z=\text{Diag(1,0,0,-1)}$. It is assumed that Alice is performing a homodyne (in the $x$ or $p$ quadrature) or heterodyne (in the $x$ and $p$ quadratures simultaneously using two homodyne detectors on the signal, split on a balanced beam splitter) measurement on one of the modes, which conditionally prepares the squeezed state with variance $1/V$ or coherent state in the other mode, respectively. The unmeasured mode is then being sent through the channel and the side channels. Such a scheme is then equivalent to a PM scheme based on squeezed or coherent states with $V_S=1/V$ or $V_S=1$, respectively (depending on the homodyne or heterodyne measurement applied by Alice) and $V_M=V-V_S$. The mode interactions in the side channels and the main channel based on the liner coupling are taken into account in the covariance matrices using the input-output relations for the quadrature vectors
$r_i=(x_i,p_i)^T$ of interacting modes 1 and 2 in the form
\begin{eqnarray}
\label{inout}
\left( \begin{array}{c}
r_1 \\
r_2
\end{array} \right)_{out} =
\left( \begin{array}{cc}
\sqrt{T}\mathbb{I} & \sqrt{1-T}\mathbb{I} \\
-\sqrt{1-T}\mathbb{I} & \sqrt{T}\mathbb{I}
\end{array} \right)
\left( \begin{array}{c}
r_1 \\
r_2
\end{array} \right)_{in},
\end{eqnarray}
where $T$ stands for the transmittance of a coupling beam splitter. Such transformations lead to changes of variances and covariances that form the resulting covariance matrices. The lower bound on the key rate secure against collective attacks is then calculated numerically using \ref{eq:von_Neumann} and \ref{eq:BosonicEF}. In the case when modulation $V_M$ is independent of the signal squeezing $V_S$ the more general entanglement-based scheme \cite{Usenko11} is used instead of the standard EPR-based purification described above.

The purification method allows us to analyze the security of the protocols in the conditions of untrusted noise by estimating the lower bound on the secure key rate and in particular to study the region of insecurity where the lower bound turns to zero. Further, we describe the theoretical purification schemes used to calculate the Holevo bound in the particular cases. Note that the purification schemes give also the same mutual information $I_{AB}$ as in the PM versions of the protocols. We also cross-check our results using the entangling cloner \cite{pm_epr} collective attack being the particular purification of the channel noise by an EPR source possessed by Eve, which is also widely used in CV QKD security analysis as a typical collective attack (see, e.g., \cite{QKDClassLim,QKDClassLim2}). The results obtained using the entangling cloner exactly confirm our calculations based on the purification models.
\subsection{Side-channel loss on the sender side}
In the case of the type-A side channel, the variance of Alice's data is unchanged and remains $V_M$ and the correlation between Alice and Bob is scaled by the channel and the side channel so that $C_{AB}=\sqrt{\eta_A\eta}V_M$. The variance of the state measured by Bob in the $x$ quadrature after the side channel and the main noisy and lossy channel $V_B=\eta[\eta_A V+\epsilon-\eta_A]+1$. 

We first investigate the influence of the side channel for the case of individual attacks with pure losses ($\epsilon =0$) to estimate the security region. Taking into account the above-given variances and correlations, the mutual information $I_{AB}$ can be directly obtained as (\ref{typeAmi}). In the case of individual attacks in the purely lossy channel, Eve is able to measure the output mode of the side channel, which we denote by $S_A$, and the output of the main channel which we denote by $E$. Therefore, the mutual information $I_{BE}$ using the symmetry of the mutual information (\ref{ixy}) is to be calculated as
\begin{equation}
I_{BE}=\frac{1}{2}\log_2\frac{V_B}{V_{B|ES_A}},
\end{equation}
where $V_{B|ES_A}$ 
is the variance of Bob's measurement conditioned by measurements of Eve on the modes $E$ and $S_A$. 
The calculations taking into account the variances of Eve's modes $V_{E}=(\eta_A V+1-\eta_A)(1-\eta)+\eta$, and 
$V_{S_A}=\eta_A+(1-\eta_A)V$ 
and correlations $C_{BE}=\eta_A\sqrt{\eta(1-\eta)}(1-V)$ 
and $C_{BS_A}=\sqrt{\eta_A\eta(1-\eta_A)}(1-V)$ result in the expression
\begin{equation}
V_{B|ES_A}=\frac{V}{\eta_A\eta(1-V)+V}
\end{equation}
from which the expression (\ref{typeAmiE}) is obtained.

In the case when the channel noise is present we model Eve's individual attack as an optimal entangling cloner \cite{pm_epr}, i.e., we assume that Eve possesses the two-mode entangled source $E_1E_2$ with the variance $N=1+\frac{\eta\epsilon}{1-\eta}$ so that the mode $E_1$ interacts with the signal and introduces the loss $\eta$ and the excess noise $\epsilon$. Eve is then able to measure three modes: the output of the side channel $S_A$ and the modes $E_1$ and $E_2$ of the entangling cloner. Therefore, the mutual information $I_{BE}$ between Eve and Bob should read
\begin{equation}
I_{BE}=\frac{1}{2}\log_2{
\frac{V_B}{
V_{B|S_AE_1E_2}
}
},
\end{equation}
where $V_{B|S_AE_1E_2}$ is the variance of Bob's measurement conditioned by measurements of Eve on the modes $S_A$, $E_1$, and $E_2$. 
The variances of the modes after the side channel and the main channel are $V_B=\eta[\eta_A V+1-\eta_A+\epsilon]+1$ 
[which also changes the mutual information (\ref{typeAmiE})], 
$V_{S_A}=\eta_A+(1-\eta_A)V$ is unchanged by the channel noise, 
$E_1=\eta N+(1-\eta)(\eta_A V+1-\eta_A)$ and $E_2=N$. 
The correlations are $C_{BS_A}=\sqrt{\eta_A\eta(1-\eta_A)}(1-V)$, 
$C_{BE_1}=\sqrt{\eta(1-\eta)}(N-\eta_A V-1+\eta_A)$, 
and $C_{BE_2}=\sqrt{(1-\eta)(N^2-1)}$. From this the conditional variance 
\begin{equation}
V_{B|S_AE_1E_2}=\frac{1+\eta_A(V-1)}{1+\eta\epsilon+\eta_A(V-1)[1-\eta(1-\epsilon)]}
\end{equation}
can be obtained and used to calculate the key rate secure against the individual attacks in a noisy channel.

In the case of collective attacks in a noisy channel and no additional manipulation on the side-channel input the security is calculated through the 4x4 covariance matrix
\begin{equation}
\gamma_{AB}=
\left(
\begin{array}{cc}
 \text{V$\mathbb{I}$} & \sqrt{\text{$\eta_A \eta $} \left(V^2-1\right)} \sigma _z \\
 \sqrt{\text{$\eta_A \eta $} \left(V^2-1\right)} \sigma _z & [(V-1) \eta_A \eta +\epsilon  \eta +1] \mathbb{I} \\
\end{array}
\right),
\label{CovM1}
\end{equation}
which describes the state shared between the trusted parties in the EPR-based version of the protocols. The conditional matrix after the measurement at Bob in particular contains $\eta_A$ separately from $\eta$ and reads
\begin{equation}
\gamma_{A|B}=
\left(
\begin{array}{cc}
V-\frac{\eta_A \eta(V^2-1)}{1+\eta(\eta_A V-\eta_A+\epsilon)} & 0 \\
 0 & V \\
\end{array}
\right).
\label{CovM2}
\end{equation}
From these two matrices the security of the protocol can be analyzed for the case of collective attacks. We do not provide the explicit expressions for the multimode covariance matrices in the further analysis since they are too lengthy; Thowever, they can be directly obtained using the input-output relations \ref{inout} and the details of the purification schemes given below. Further, we present the purification schemes for the different methods of the side-channel decoupling as well as the changes of the variances and correlations of measured data upon the additional manipulations on the side channels. 

\begin{figure}
	\centering
		\includegraphics[scale=0.55]{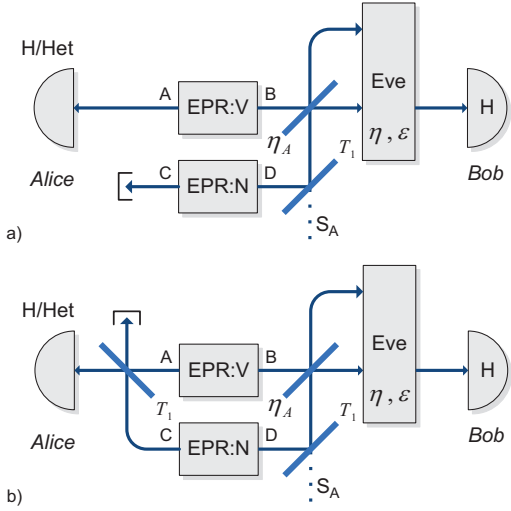}
	\caption{(a) Theoretical purification of the equivalent PM scheme with the side-channel loss on the sender side and thermal noise applied on the input of the side channel. (b) Purification of the PM scheme with the side-channel loss on the sender side and known modulation applied on the input of the side channel. Homodyne or heterodyne detection is applied at the side of Alice in both the cases depending on the protocol.}
	\label{scApur12}
\end{figure}

First, if the uncorrelated noise is added to the input of the side channel, it is modeled as the coupling of one of the modes of the EPR source N [see Fig. \ref{scApur12}, (a)] to the side-channel input using a strongly unbalanced beam splitter with transmittance (for mode $S_A$) $T_1 \to 1$. The variance of the source is respectively set to $N=V_{NS}/(1-T_1)$; this way the noise is added losslessly. 

The state of the modes ABCD contains all the purified trusted noise and the Holevo bound for the standard Gaussian protocols is then calculated following the purification method as $\chi_{BE}=S(\gamma_{ABCD})-S(\gamma_{ACD|B})$. If the known modulation is applied to the side-channel input, then the purification is based on a similar scheme but the second mode of the EPR source N is coupled to the mode A, measured by Alice. This way the displacement that is applied to the input of the side channel is also added to the displacement measured by Alice [see Fig. \ref{scApur12}, (b)]. Alice's data in this case have the variance $V_A=V_M+V_{NM}$, while the correlation with Bob after the side channel and the main channel is given by $C_{AB}=\sqrt{\eta}(\sqrt{\eta_A}V_M+\sqrt{1-\eta_A}V_{NM})$. Bob's measured variance is $V_B=\eta[\eta_A(V-V_{NM})-\eta_A+\epsilon]+1$. From this expression the expression for the mutual information (\ref{typeAmi2}) is directly obtained.

The calculations are then similar to the previous case. In both cases, if the generalized scheme in which modulation is independent from the signal states is to be used, then the main source EPR:V is replaced with the respective entanglement-based generalized preparation as described in \cite{Usenko11}. The manipulations on the side channel remain purified as described above.

Finally, if the correlated displacement is added and the input of the side channel is additionally squeezed to $V_S$, then the variance of Alice's data remains $V_M$, but the correlation with Bob is changed to $C_{AB}=\sqrt{\eta}(\sqrt{\eta_A}+k\sqrt{1-\eta_A})V_M$ and the variance of the state measured at the Bob's side is $V_B=\eta[2kV_M\sqrt{\eta_A(1-\eta_A)}+k^2V_M(1-\eta_A)-\eta_A V_M+V_S+\epsilon-1]+1$. From this the expression for the mutual information can be obtained in the general case. 

\begin{figure}
	\centering
		\includegraphics[scale=0.55]{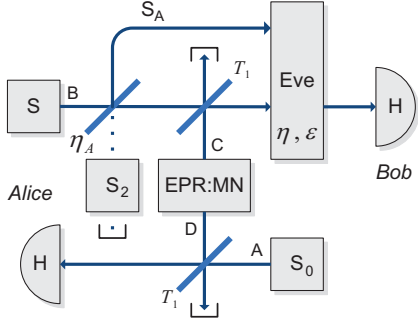}
	\caption{Theoretical purification of the equivalent PM scheme with the side-channel loss on the sender side and the optimal correlated modulation applied on the input of the side channel precessed by the optional squeezed-state preparation on the source $S_2$ in the case of the squeezed-state protocol as depicted in Fig. \ref{scheme-corr}. The source $S_0$ produces an infinitely squeezed state, the entangled source EPR:MN provides the modulation of the signal, and the trusted parties perform homodyne detection on their respective modes A and B.}
	\label{scApur3}
\end{figure}

For the optimal $k=\sqrt{(1-\eta_A)/\eta_A}$ and with the side-channel input substituted by the squeezed state with variances $V_S$ and $1/V_S$ (in the case of the squeezed-state protocol) it is easy to see that the main signal mode and the side-channel output described by $x'_A=x_S\sqrt{\eta_A}+x_{SCA}\sqrt{1-\eta_A}+x_M(\sqrt{\eta_A}+k\sqrt{1-\eta_A})$ and $\tilde{x}'_{SCA}=x_{SCA}\sqrt{\eta_A}-x_S\sqrt{1-\eta_A}+x_M(k\sqrt{\eta_A}-\sqrt{1-\eta_A})$, respectively become completely uncorrelated, i.e., $Cov(x'_A,\tilde{x}'_{SCA})=0$. Therefore, the side channel becomes decoupled from the main signal. 
For the calculations of the Holevo bound the equivalent scheme depicted in Fig. \ref{scheme-corr} must be purified. This is done by introducing the EPR source MN (see Fig. \ref{scApur3}) with variance $V_M/[\eta_A(1-T_1)]$. It is coupled to the signal state produced by the source $S$ in the mode $B$ and to the infinitely squeezed state used for simulating the detection, produced by the source $S_0$ in the mode $A$ on the strongly unbalanced beam splitters with the transmittance for the modes $A$ and $B$ being $T_1 \to 1$. The input of the side channel is optionally squeezed using squeezer $S_2$ in the case of the squeezed-state protocol. The state that is then sent through the channel (mode B) is then defined by the single-mode covariance matrix $\gamma_B=\text{Diag}(T_1 V_S+V_M/\eta_A,T_1/V_S+V_M/\eta_A)$. In the limit $T_1 \to 1$ this is equivalent to the preparation of a signal state $V_S$, attacked by the side channel with input $V_S$ (prepared by Alice in the case of the squeezed-state protocol), which leaves the signal state unchanged, and subsequent symmetrical (having the same variance in the both the $x$ and $p$ quadratures) Gaussian modulation with variance $V_M/\eta_A$. The correlation between the measurements at Alice and at Bob in the absence of the quantum channel is $C_{AB}=-\sqrt{(V_M/\eta_A)^2-(1-T_1)^2}$, which is equivalent to the modulation with variance $V_M/\eta_A$ applied by Alice in the PM setup when $T_1 = 1$. The state that is measured by Alice (mode A) is defined by $\gamma_A=\text{Diag}(T_1 V_0+V_M/\eta_A,T_1/V_0+V_M/\eta_A)$, where $V_0$ is the variance of the squeezed state, produced by the source $S_2$. The first element of matrix $\gamma_A$, which is measured by the $x$-quadrature homodyne measurement, in the limit of $V_0 \to 0$, corresponds to Alice perfectly knowing the displacements of the modulation $V_M/\eta_A$ in the PM setup. After the measurement at the Alice's side the state that is conditionally prepared on the channel input is given by $\gamma_{B|A}=\text{Diag}\big(T_1 V_S+[\eta_A(T_1-1)^2+T_1V_0V_M]/(\eta_A T_1 V_0+V_M),T_1/V_S+V_M/\eta_A\big)$, which in the regime of $T_1=1$ and $V_0=0$ gives $\text{Diag}(V_S,1/V_S+V_M/\eta_A)$, corresponding to the modulation of the signal state $\text{Diag}(V_S,1/V_S)$ with variance $V_M/\eta_A$ in both the quadratures with only one value $x$ being kept. Our purification scheme (see Fig. \ref{scApur3}) is therefore equivalent to the PM one (shown in Fig. \ref{scheme-corr}), providing (in the limits $T_1 \to 1$ and $V_0 \to 0$) the same variances and correlations, and resulting in the same conditional states. Moreover, the developed scheme allows purification of practically any PM scheme being more adjustable than the standard EPR-based approach \cite{Lod}. The asymmetrical modulation can be introduced by the general preparation of the state $EPR:MN$ using two different orthogonally squeezed states, however, such an extension was not needed in the tasks of the present paper. 

In the purification scheme the state of the modes $ABCD$ is pure, while the channel noise and loss introduce impurity to the state. The mode $S_A$ is not relevant in the scheme since it is uncorrelated from the rest of the setup due to the equality of the variances of modes $A$ and $S_A$ prior to the side-channel coupling $\eta_A$ (it is shown on the scheme only for explanatory purposes). Then the Holevo bound is calculated as $\chi_{BE}=S(\gamma_{ABCD})-S(\gamma_{ACD|B})$.
\subsection{Side-channel noise infusion on the receiver side} 
In the case of the type-B side channel the variance of Alice's data remains $V_M$ and the 
correlation between Alice and Bob is scaled by the channel and the side channel so that $C_{AB}=\sqrt{\eta\eta_B}V_M$. The variance of the state measured by Bob in the $x$ quadrature after the main noisy and lossy channel and the side channel is $V_B=\eta_B[\eta (V+\epsilon)+1-\eta]+(1-\eta_B)V_N$. 

Let us first consider the individual attacks in the purely attenuating main channel, i.e., $\epsilon=0$. Taking into account the above-given variances and correlations, the mutual information $I_{AB}$ can be directly obtained as (\ref{typeBmi}). In the case of individual attacks in the purely lossy channel Eve is able to measure the output mode of the main channel, which we denote by $E$. Moreover, Eve controls the input of the noisy side channel, which we introduce as an entangling cloner attack, which was shown to be optimal in the case of individual attacks \cite{pm_epr}. Therefore, we assume that Eve possesses the two-mode entangled source $E_1E_2$ with the variance $V_N$ and is able to measure one of the modes $E_1$, while the other mode $E_2$ is coupled to the signal on the $\eta_B$ beam splitter. Therefore, the mutual information $I_{BE}$ using the symmetry of the mutual information (\ref{ixy}) is to be calculated as
\begin{equation}
I_{BE}=\frac{1}{2}\log_2{\frac{V_B}{V_{B|EE_1}}},
\end{equation}
where $V_{B|EE_1}$ is the variance of Bob's measurement conditioned by measurements of Eve on the modes $E$ and $E_1$. The calculations taking into account the variances of Eve's modes $V_{E}=V(1-\eta)+\eta$, and $V_{E_1}=V_N$ (since is Eve is measuring the mode of the cloner that did not interact with the signal) and correlations $C_{BE}=\sqrt{\eta\eta_B(1-\eta)}(1-V)$ and $C_{BE_1}=\sqrt{(1-\eta_B)(V_N^2-1)}$ (the latter provided by the correlations within the entangling cloner) result in the expression
\begin{equation}
V_{B|EE_1}=\frac{\eta_B V}{V(1-\eta)+\eta}+\frac{1-\eta_B}{V_N}
\end{equation}
from which the expression (\ref{typeBmiE}) is obtained.
\begin{figure}
	\centering
		\includegraphics[scale=0.55]{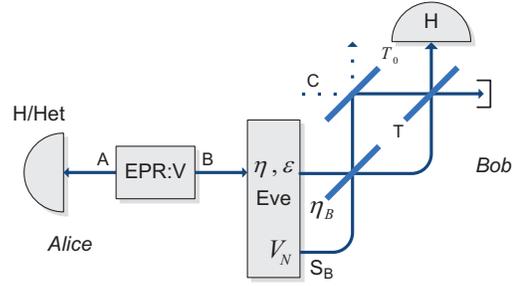}
	\caption{Theoretical purification of the equivalent PM scheme with the side-channel noise addition on the receiver-side and the optimal monitoring of the side-channel output, represented by the interferometric scheme applied on the output of the side channel prior to the trusted detection; the homodyne or heterodyne detection is applied at the side of Alice depending on the protocol.}
	\label{scBpur}
\end{figure}

If the main homodyne detector $H$ and the monitoring detector $H'$ (see Fig. \ref{solution}) are both imperfect with loss $\eta_D$ and noise of the variance $V_D$, which is coupled to the signal with the ratio $\eta_D$ \cite{Lod}, then the quadratures measured by the detectors $H$ and $H'$ will be given by
\begin{equation}
x_B'=\sqrt{\eta_D}(x_B\sqrt{\eta_B}+x_{SCB}\sqrt{1-\eta_B})+x_1\sqrt{1-\eta_D}
\end{equation}
and
\begin{equation}
x_{SCB}'=\sqrt{\eta_D}(-x_B\sqrt{1-\eta_B}+x_{SCB}\sqrt{\eta_B})+x_2\sqrt{1-\eta_D}
\end{equation}
respectively, where $x_1$ and $x_2$ are the quadrature values associated with the detector noise such that $Var(x_1)=Var(x_2)\equiv V_D$. The weighted difference $\Delta x=gx_B'-g'x_{SCB}'$ will then be given by
\begin{widetext}
\begin{equation}
\Delta x=x_B\sqrt{\eta_D}(g\sqrt{\eta_B}+g'\sqrt{1-\eta_B})+x_{SCB}\sqrt{\eta_D}(g\sqrt{1-\eta_B}-g'\sqrt{\eta_B})+\sqrt{1-\eta_D}(gx_1-g'x_2)
\end{equation}
\end{widetext}
By setting the weights of the difference to $g=\sqrt{\eta_B}$ and $g'=\sqrt{1-\eta_B}$ the result of the subtraction becomes
\begin{equation}
\Delta x=x_B\sqrt{\eta_D}+\sqrt{1-\eta_D}(x_1\sqrt{\eta_B}-x_2\sqrt{1-\eta_B}),
\end{equation}
where the noise of the side channel given by the quadrature value $x_{SCB}$ is {\it completely} removed. The variance of the weighted difference then becomes $Var(\Delta x)=\eta_D V_B+(1-\eta_D)V_D$, i.e. equivalent to the measurement of the signal $x_B$ on the imperfect homodyne detector with loss $\eta_D$ and noise $V_D$; the scaling $\sqrt{\eta_D}$ then also applies to the correlation $C_{AB}$. When the detection is purely lossy, i.e., $V_D=1$, the expression then further simplifies as $Var(\Delta x)=\eta_D V_B+1-\eta_D$.

In the case of collective attacks in the noisy channel and in the presence of the type-B side channel we use the purification scheme based on the entangled source of modes A and B of variance V with mode A measured on the Alice's side with the homodyne or heterodyne detector. In this case the noisy mode $S_B$ is assumed to be purified by Eve (see Fig. \ref{scBpur}). However, it is reflected by the beam splitter with transmittance $T_0=0$ fed by the vacuum input and the fully reflected mode C is then coupled on the unbalanced beam splitter $T$ with the signal mode B. Then all the impurity of the state shared between Alice and Bob is attributed to Eve and the following equalities hold: $S(\gamma_E)=S(\gamma_{ABC})$ and $S(\gamma_{E|B})=S(\gamma_{AC|B})$. 

Further, we equivalently represent the type-B side-channel output monitoring and data manipulation by an interferometric scheme, when the outputs of the side-channel coupling $\eta_B$ (modes $B$ and $C$ in the purification scheme) are coupled again on a beam splitter with transmittance $T$. The case when the interferometric setup is properly balanced, i.e., $T=\eta_B$, corresponds to the optimized monitoring on the output of the type-B side channel, as described in the main text. Indeed, the quadrature measured on the signal mode $B$ after all the interactions is given by $x'_B=x_B[\sqrt{T\eta_B}+\sqrt{(1-T)(1-\eta_B)}]+x_{SCB}[\sqrt{\eta_B(1-T)}-\sqrt{T(1-\eta_B)}]$, where $x_B$ is the main signal mode before the side-channel interaction and $x_{SCB}$ is the side-channel input prior to interaction. It is easy to see that upon $T=\eta_B$ the resulting quadrature $x'_B=x_B$ of mode $B$ contains no side-channel noise due to the destructive interference and the negative effect of the side channel is removed completely. Note that if the reflection of the mode $S_B$ on the beam splitter $T_0$ would be absent and the mode $B$ would be directly coupled to the mode $S_B$ on the beam splitter with transmittance $T$, the equivalent measurement that removes the type-B side channel would be on the mode $S_B$ upon $T=1-\eta_B$. The described scheme allows calculations using the purification method simply as $\chi_{BE}=S(\gamma_{ABC})-S(\gamma_{AC|B})$ since the side-channel output monitoring emulated by the interferometric setup does not change the purity of the states. The performance of the protocol thus becomes equivalent to the one of the protocol without the type-B side channel, which is confirmed in the case of collective attacks in a noisy channel. In the case of the generalized preparation (when modulation is independent of the signal state variance) we apply a similar scheme but replace the EPV:V source with the generalized entangled state preparation as described in \cite{Usenko11}.

In the case of the interferometric-type interaction between the signal and the type-B side channel, as shown in Fig. \ref{KR-B-adv}, the mode transformations during the interactions become more complex and read
\begin{widetext}
\begin{eqnarray}
\lefteqn{x_B'=x_B\big[\sqrt{\eta_{B1}\eta_{B2}}-\cos{\phi}\sqrt{(1-\eta_{B1})(1-\eta_{B2})}\big]+x_{SCB}\big[\sqrt{\eta_{B2}(1-\eta_{B1})}+{}} \nonumber \\
& {}+\cos{\phi}\sqrt{\eta_{B1}(1-\eta_{B2})}\big]-p_B\sin{\phi}\sqrt{(1-\eta_{B1})(1-\eta_{B2})}+p_{SCB}\sin{\phi}\sqrt{\eta_{B1}(1-\eta_{B2})}
\end{eqnarray}
and
\begin{eqnarray}
\lefteqn{x_{SCB}'=x_B\big[-\sqrt{\eta_{B1}(1-\eta_{B2})}-\cos{\phi}\sqrt{\eta_{B2}(1-\eta_{B1})}\big]+
x_{SCB}\big[\cos{\phi}\sqrt{\eta_{B1}\eta_{B2}}-{}} \nonumber \\
& {}-\sqrt{(1-\eta_{B1})(1-\eta_{B2})}\big]-
p_B\sin{\phi}\sqrt{\eta_{B2}(1-\eta_{B1})}+
p_{SCB}\sin{\phi}\sqrt{\eta_{B1}\eta_{B2}}
\end{eqnarray}
\end{widetext}
now involving the contributions from the $p$ quadratures $p_B$ and $p_{SCN}$ of the signal and side-channel noise modes respectively, which is caused by the phase shift in the interaction. This prevents the complete decoupling of the type-B side channel by simple manipulation on the homodyne measurement results in the form $gx_B'-g'x_{SCB}'$, as illustrated in Fig \ref{KR-B-adv} (plotted based on the numerical calculations using the equivalent transmittance $T$ in the purification-based scheme). The complete decoupling in such a case is possible only when $\phi=0$ and the cross-quadrature terms are absent. 
\end{document}